\documentclass[prd,preprint,eqsecnum,nofootinbib,amsmath,amssymb,
               tightenlines,floatfix]{revtex4}
\input epsf
\usepackage{bm}
\def\putbox#1#2{\epsfxsize=#1\textwidth\epsfbox{#2}}
\def\centerbox#1#2{\centerline{\epsfxsize=#1\textwidth\epsfbox{#2}}}

\def\be{\begin{equation}}
\def\ee{\end{equation}}
\def\bea{\begin{eqnarray}}
\def\eea{\end{eqnarray}}
\def\sss{\scriptscriptstyle}

\def\Eq#1{Eq.~(\ref{#1})}
\def\Tc{T_{\rm c}}

\def\OO{{\cal O}}

\def\mH{m_{\sss H}}
\def\mS{m_{\sss S}}
\def\lH{\lambda_{\sss H}}
\def\lS{\lambda_{\sss S}}
\def\Vvac{{V_{\rm vac}}}
\def\VT{V_{\sss T}}

\def\p{{\bm p}}

\def\nb{f_{\sss \rm B}}

\def\half{{\textstyle{\frac 12}}}

\def\lsim{\mbox{~{\raisebox{0.4ex}{$<$}}\hspace{-1.1em}
        {\raisebox{-0.6ex}{$\sim$}}~}}

\advance\parskip 2pt

\begin{document}

\title
    {
      Can electroweak bubble walls run away?
    }

\author{Dietrich B\"odeker}
\affiliation{
Fakult\"at f\"ur Physik,
Universit\"at Bielefeld,
D-33501 Bielefeld,
Germany
}

\author{Guy D.~Moore}
\affiliation
    {%
    Department of Physics,
    McGill University, 
    3600 rue University,
    Montr\'{e}al, QC H3A 2T8, Canada
    and \\
Fakult\"at f\"ur Physik,
Universit\"at Bielefeld,
D-33501 Bielefeld,
Germany
    }%

\date {March 2009}

\begin {abstract}%
    {%
In extensions of the Standard Model with SU(2) singlet scalar fields, there
can be regions of parameter space for which the electroweak phase transition
is first order already at the mean-field level of analysis.  We show that in
this case the phase interface (bubble wall) can become ultra-relativistic,
with the relativistic gamma factor $\gamma = (1-v_{\rm wall}^2)^{-1/2}$
growing linearly with the wall's propagation distance.  We provide a simple
criterion for determining whether the bubble wall ``runs away'' in this way or
if $\gamma$ approaches a terminal value.
    }%
\end {abstract}

\maketitle
\thispagestyle{empty}

\section {Introduction}

Electroweak symmetry breaking is a central feature of the Standard Model,
arising because the Lagrangian mass squared for the Higgs field is negative.
However, interaction with a thermal medium rather generically raises the
effective (thermal) mass squared of scalar particles, leading to high
temperature symmetry restoration \cite{T_symm,DolanJackiw}.  Such symmetry
restoration is expected in the Standard Model at temperatures of order
$T \sim \mH \sim 100$GeV, which presumably occurred very early in the hot Big
Bang.

Such symmetry breaking in a gauge-singlet scalar field would automatically
imply a thermal phase transition between symmetry-restored (high temperature)
and symmetry broken (low temperature) phases.%
\footnote{
     It is possible in some models to have thermal effects induce symmetry
     breaking, rather than restoration \cite{Weinberg}.  But this requires
     rather special coupling relations and is not important to the rest of our
     discussion.
     }
But the Higgs field (or fields) transforms nontrivially under a nonabelian
gauge symmetry.  The notions of vacuum expectation value and symmetry breaking for
such a scalar are only defined perturbatively, and there is no guarantee that
there will be a phase transition associated with electroweak symmetry breaking
\cite{no_transition}.  In particular, within the Minimal Standard Model at
allowed values for the Higgs mass, no electroweak phase transition exists
\cite{Rummukainenetal}.  But an electroweak phase transition would occur in
numerous phenomenologically interesting extensions of the Standard Model.
Such a transition is interesting cosmologically for a number of reasons:
\begin{itemize}
\item
A first order electroweak phase transition might lead to electroweak
baryogenesis \cite{original,Baryo_reviews}, giving rise to the baryon number
observed in the universe today.
\item A first order phase transition might produce primordial magnetic fields
  \cite{TurnerWidrow,Baym} which might seed the formation of today's galactic
and extragalactic magnetic fields.
\item
A first order electroweak phase transition might give rise to observable
gravitational wave backgrounds \cite{hogan,Kosowsky,EW_gravwave} in interesting
frequency ranges for future experiments (around $10^{-3}$ Hertz).
\end{itemize}

There are two key attributes of the transition which are important in these
scenarios.  One is the strength of the electroweak phase
transition (for instance, the available free energy and latent heat as a
fraction of the energy density of the electroweak plasma).  This is determined
by model parameters (field content, masses and
coupling constants) which are not known.  However we have developed the
theoretical tools for determining the strength of the transition, given the
model parameters \cite{Kajantieetal}.  The other attribute is the bubble wall
velocity, meaning the propagation speed of the interface between high
temperature and low temperature phases, while the transition is taking place.
This velocity is relevant to all physical consequences of the phase
transition.  In particular:
\begin{itemize}
\item
Electroweak baryogenesis scenarios seem to be quite sensitive to the bubble
wall velocity \cite{CohenKaplanNelson,Tomislav,Carena,Schmidt,Cline}.  A slow
bubble wall allows nonequilibrium conditions to exist ``in front'' of
the bubble wall, in the symmetric electroweak phase where baryon number
violating processes are fast \cite{CohenKaplanNelson}.
A thin, very fast bubble wall would provide a nonequilibrium environment only
in the broken phase, probably making baryogenesis very inefficient.
\item
Gravitational waves are of maximum strength if the bubble walls are
highly relativistic.  Slow subsonic bubble walls give rise to rather
extended and smooth hydrodynamic waves, which should be poor sources of
gravitational waves (see \cite{Kurki-Suonio} for the
hydrodynamics of transitions with finite bubble wall velocities).
On the other hand, if the bubble walls propagate ultra-relativistically
and a nontrivial fraction of the free energy density of the plasma goes
into their acceleration, it could give rise to very interesting
gravitational wave signals.%
\footnote{
    Steinhardt has argued \cite{Steinhardt} that bubble walls cannot
    approach the speed of light if there is a plasma present.  However
    this result was based on assuming that the Chapman-Jouguet condition
    is fulfilled, that is, that the plasma behind the wall moves at the
    speed of sound with respect to the wall.  This assumption makes
    sense for burning detonations, where the (chemical) free energy is
    released behind rather than at the shock, in a region thick enough
    to apply hydrodynamics; then the fluid must be subsonic in order to
    push the shock.  But, as carefully demonstrated by Laine \cite{Mikko},
    this assumption is not valid for cosmological phase
    transitions, where the free energy is liberated internally in the
    wall; as a result more general detonation solutions are possible.
    We complete Laine's treatment in the Appendix, where we show that
    runaway walls are thermodynamically consistent under quite weak
    conditions.
    }
\end{itemize}
In a first-order cosmological phase transition in vacuum \cite{coleman},
the energy density of the false vacuum is completely converted
into kinetic energy of the bubble walls. As a result, the wall velocity keeps
increasing and approaches the speed of light ($\gamma\equiv 1/\sqrt{1-v^2}$
grows linearly with the distance a bubble wall propagates).  Either this
behavior, or highly relativistic walls but with a terminal $\gamma$
factor, have often been assumed in the gravitational wave literature
\cite{Kosowsky}.  On the contrary, the electroweak baryogenesis literature has
generally assumed that interactions with the plasma impede the bubble wall and
keep its velocity subsonic, $v < 1/\sqrt{3}$, as the
rather limited literature computing the wall velocity suggests
$v<0.2$ \cite{Turok,Dine,Khlebnikov,Arnold,MooreProkopec,John,Moore}.

The goal of this paper is to determine under what situations the electroweak
bubble wall can ``run away'' and propagate with macroscopic gamma factor
$\gamma \equiv (1-v^2)^{-1/2} \gg 1$, and when it more modestly propagates
with $\gamma v \lsim 1$.  At the thermodynamic level, we show in
Appendix \ref{the_appendix} that runaway walls are logically possible
under quite weak conditions; but whether they actually occur depends on
how much entropy is produced by the passing wall, which is a problem in
the microscopic wall dynamics.  However, the relevant
physics is actually quite
simple, and the right tools already exist in the literature.  This paper could
have been written more than 10 years ago.  However we find that the results
neither appear in the literature nor are well known by either community.

Generally, at the mean field level models without electroweak SU(2) singlet
scalar fields predict a second order electroweak phase transition.  The
transition in such a theory {\em may} be a so-called ``fluctuation induced''
first order transition (as for the Standard Model with an unphysically light
Higgs field).  We will show that the bubble wall never runs away in such
transitions.  On the other hand, theories with SU(2) singlet scalars can
in some cases provide a first-order electroweak (SU(2) breaking) phase
transition already at the mean-field level of analysis.  This is true
for instance in the nMSSM \cite{concrete} (though not for the MSSM).
Indeed, as we discuss below, it is true already for the simplest
extension of the Standard Model, a theory with an additional real
singlet scalar. We will show the such ``mean-field first
order'' electroweak transitions generically, but not invariably, have
``runaway'' bubble walls, with $\gamma$ growing linearly with wall propagation
distance.  We present a simple criterion to determine whether or not such
``runaway'' will occur.

Specifically, the phase structure is determined by an effective
potential $V(h)$ which is the sum of a vacuum part $\Vvac(h)$ and a thermal
contribution $\VT(h)$.  The transition occurs when $V(h)=\Vvac+\VT$ has two
local minima, a shallower ``symmetric'' $s$-minimum and a deeper
``broken'' $h$-minimum; the
exact criterion for how much deeper the $h$ minimum need be is
determined by a bubble nucleation calculation \cite{bubble_nuc}.
If replacing $\VT(h)$ with its 2-nd
order Taylor approximation%
\footnote{When multiple scalar fields change VEVs at the transition, this
  expression generalizes to the quadratic form 
  $\half \sum_{ij} s_i s_j d^2 V/ds_i ds_j$, 
  where $s_i$ runs over all scalars which change VEVs.}
$(h-h_{\rm sym})^2 \VT''(h_{\rm sym})/2$
removes the ``broken'' minimum or raises it above
the symmetric one, the bubble wall cannot run away.  If, after this
replacement, the ``broken'' minimum remains deeper than the symmetric one,
then the bubble wall will run away.  The rest of the
paper derives and clarifies this result.

The paper is organized as follows.  In Section \ref{sec:1looppot} we review
briefly how the 1-loop thermal effective potential arises.  Section
\ref{sec:fric} gives a simple derivation of the friction on the electroweak
bubble wall when it is propagating ultra-relativistically with respect to the
plasma.  Section \ref{sec:analysis} illustrates our result using a
specific model (a toy version of the nMSSM), showing how to determine whether
or not runaway occurs as a function of model parameters.
Section \ref{sec:conclusions} presents some conclusions.
There is also an appendix, which discusses the thermodynamic criteria
under which runaway walls are consistent.

\section{Thermal effective potential:  a review}
\label{sec:1looppot}

Our arguments will be based on the way that the thermal effective
potential arises from the particles making up the thermal bath.
Therefore we start by reviewing this physics.
Consider for simplicity a theory of two real scalar fields $h,s$
with tree-level effective potential
\be
\Vvac (h,s) = \frac{\mH^2}{2} h^2 + \frac{\mS^2}{2} s^2
   + \frac{\lH}{4} h^4
   + \frac{\lS}{4} s^4
   + \frac{\lambda}{4} h^2 s^2 \,.
\ee
At zero temperature we determine the vacuum configuration by seeking the global
minimum of the potential with respect to $h,s$.  If,
for instance, $\mH^2$ is negative but $\mS^2$ and $\lambda$ are positive, the
global minimum will be at $h_0 = \sqrt{-\mH^2/\lH}$ and $s_0=0$.  In this toy
model such a nonzero value spontaneously breaks the 
${\bf Z}_2 \times {\bf Z}_2$ symmetry group down to ${\bf Z}_2$.

At finite temperature we must take into account, besides the energy density
arising from $\Vvac(h,s)$, also the energy density of thermal
excitations of these (and any other) fields.  Fortunately one may define a
thermal contribution to the potential $\VT(h,s)$ describing the free energy
density of the plasma.  The total potential $V = \Vvac + \VT$ describes
the total free energy as a function of $h,s$, and its
minimum describes the most favored classical background.  One efficient
way to compute the thermal contributions to $\VT(h,s)$ is
differentially (the tadpole method \cite{Weinberg});
we compute the derivatives $d\VT(h,s)/dh$ and $d\VT(h,s)/ds$, which can be
integrated to recover the full potential.  In particular,
to compute the difference between $\VT(h,s)$ and $\VT(h{+}\delta h,s)$,
calculate the change in all particle energies in changing the value of $h$:
\be
\delta \VT = \sum_{a} \int \frac{d^3 p}{(2\pi)^3}
   \nb(E_{p,h,a}) \frac{dE_{p,h,a}}{dh} \delta h
 = \sum_a \int \frac{d^3 p}{(2\pi)^3} \frac{\nb(E_{p,h,a})}{2E_{p,h,a}}
   \frac{dm^2_{a}}{dh} \delta h \,,
\label{eq:tadpole}
\ee
where the sum $\sum_a$ is over all fields (for us, just $h,s$),
$E_{p,h,a}=\sqrt{p^2+m^2_a(h,s)}$ and $m^2_a(h,s)$ is the $h,s$ dependent mass
squared for particle type $a$.  For a scalar field $m^2$ is the second
derivative of $V(h,s)$ with respect to the scalar field value; for
instance, in our toy model 
$m_s^2(h,s) = \mS^2 + 3\lambda_s s^2 + \lambda h^2/2$.

In addition to this energy cost, there is a free energy cost associated with
changing the particle occupancies $\nb(E_{p,h,a})$ from the values relevant at
field value $h$ to those relevant at field value $h{+}\delta h$.  But this
free energy cost is quadratic in $\delta h$, since the equilibrium occupancies
are a minimum of the free energy.%
\footnote{
     This is an example of the usual perturbation theory result that
     changes to the state first give rise to shifts in the (free) energy
     at second order in a perturbation.
     }
Therefore, if we treat $\delta h$ as
infinitesimal we may drop this term, and reconstruct the thermal contribution
to the effective potential by integrating \Eq{eq:tadpole} -- {\em provided} we
account for the way that $E_{p,h,a}$ and $\nb(E_{p,h,a})$ vary with
$h$ as we integrate.

Often the particle masses $m_a(h,s)$ are small compared to $\pi T$.
Then we can approximate $\nb(E)/2E$ in the second expression in
\Eq{eq:tadpole} with $\nb(p)/2p$.  In this case we can perform the
integral:
\be
\frac{d\VT}{dh} = \sum_a \frac{dm_a^2}{dh} \int \frac{d^3 p}{(2\pi)^3}
\frac{\nb(E_{p,h,a})}{2E_{p,h,a}}
\simeq \sum_a \frac{dm^2_a}{dh}
\int \frac{d^3 p}{(2\pi)^3}
\frac{\nb(p)}{2p}
= \sum_a \frac{dm^2_a}{dh} \frac{T^2}{24}
\ee
which is trivial to integrate:
\be
\VT(h) = \frac{T^2}{24} \sum_a m^2_a(h) + V_0(T)  \,,
\label{eq:meanfield1}
\ee
with $V_0(T)$ an undetermined (and for us irrelevant) integration constant.
A slightly less drastic approximation, useful if $m^2(h_1)-m^2(h_2) \ll T^2$
but valid even if $m^2$ itself is relatively large,
is to neglect the $h$ dependence of
the particle masses in the expression for $E$, replacing $E_{p,h,a}$
with $E_{p,h_1,a}$:
\be
\VT(h_2) - \VT(h_1) \simeq \sum_a \Big( m^2_a(h_2) - m^2_a(h_1) \Big)
\int \frac{d^3 p}{(2\pi)^3} \frac{\nb(E_{p,h_1,a})}{2E_{p,h_1,a}}
\,.
\label{eq:meanfield2}
\ee
This approximation is the same as determining $\VT(h_2)$ by
performing a 2'nd order Taylor series expansion of $\VT$ about $h_1$
-- provided that the potential is at most quartic, and particle squared
masses quadratic, in the condensate values of scalar fields (true at
tree level in renormalizable theories).
In what follows we will call this the mean field approximation to the
effective potential.  In our toy model, if only the $h$
field takes a nonzero background value, the effective mass for the $s$ field
is $d^2 V_{\rm vac} /ds^2 = \mS^2 + \lambda h^2/2$.  In this case the $s$ field
part of the thermal contribution to
the potential, making the stronger approximation \Eq{eq:meanfield1}, would be
$\VT(h) = T^2\mS^2/24 + T^2 \lambda h^2/48$ (plus a constant).
The first term here is also a constant; the second
looks like a correction (positive if $\lambda > 0$) to the $h$ field mass
squared.  At temperatures where $T^2 \lambda/24 > -\mH^2$, there is a
(meta)stable minimum at $h=0$ and symmetry will be restored.

\begin{figure}
\centerbox{0.6}{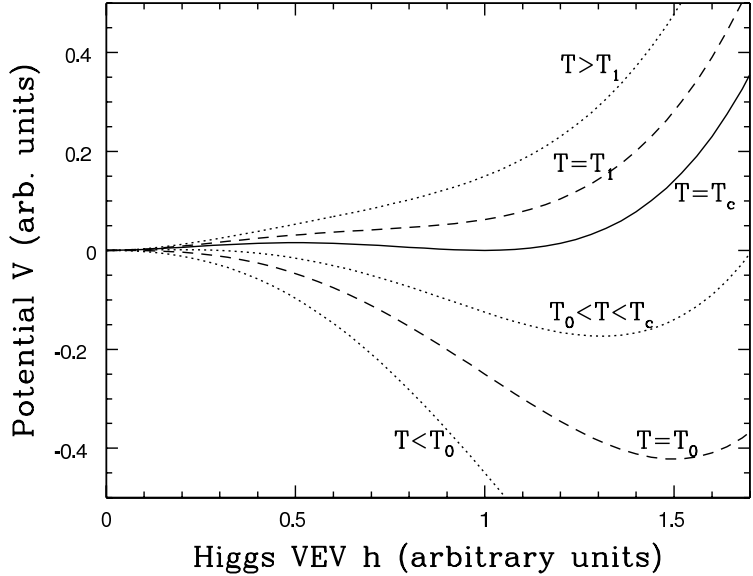}
\caption[Potential with first order transition]
{\footnotesize Example of the potential $V=\Vvac+\VT$ in our toy model,
 for parameters which give a first order transition.  The curves
 are $V(h)$ at a series of temperatures from high (top) to low (bottom);
 the dotted curves are the potential when the $h=0$ phase becomes spinodally
 unstable ($T_0$) and where the $h\neq 0$ phase becomes unstable ($T_1$).  The
 solid curve is the potential at the equilibrium temperature ($T_c$) where both
 phases have the same minimum value.
\label{fig:1order}}
\end{figure}

Often this mean-field approximation is insufficient.  For instance, in this
toy example it predicts that
the phase transition should be of second order.
Continuing to treat $m_a \ll \pi T$, but being more careful
to account for the difference between $\nb(E)/E$ and $\nb(p)/p$, one finds a
series expansion in $m_a/\pi T$:
\be
\VT(h) = \sum_a \frac{T^2 m^2_a(h)}{24}
 - \frac{T m^3_a(h)}{12\pi} + \OO(m^4/\pi^2)
\ee
which can predict a first-order phase transition, as illustrated
in Figure \ref{fig:1order}.
Since this transition is only first order because of changes in
$\nb(E)/E$ (changes in the size of the fluctuations), it is called a
fluctuation induced phase transition.

There are two potentially significant corrections to this simplistic
description.  One is that it has treated thermal particles as free, neglecting
their mutual interactions.  These give rise to higher order (``two loop'')
corrections to the thermal potential computed here.  These corrections
can be important in establishing the order and strength of a phase transition
when it is second or weakly first order.  Second, there are subtleties in
defining an effective potential if a scalar which may take on a vacuum value
is a non-singlet under gauge interactions.  Fortunately, for the electroweak
transition to be phenomenologically interesting, it must be relatively
strong.  In this case neither of these
subtleties is very important and the one-loop
treatment described here is generally adequate (unless the vacuum theory is
strongly coupled, in which case we have little to say).

\section{Friction on a relativistic bubble wall}
\label{sec:fric}

\begin{figure}
\centerbox{0.6}{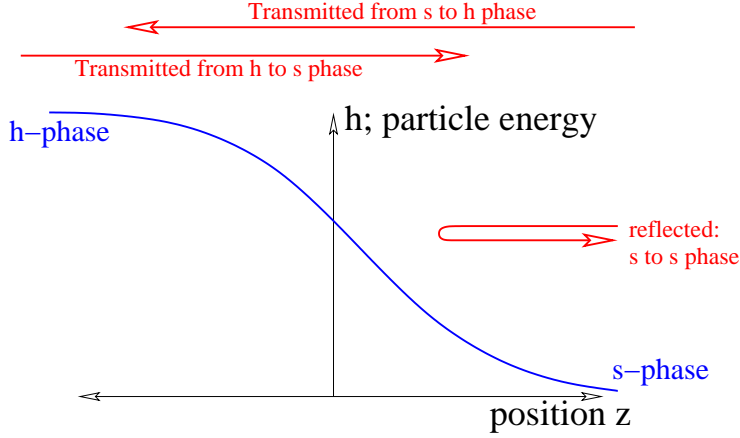}
\caption[Bubble wall illustration]
{\footnotesize Cartoon of a bubble wall:  the scalar field $h$ as a function of position
$h(z)$.  Particles ``hit'' the wall from either side, inducing net forces.
\label{fig:wall}}
\end{figure}

When an effective potential has two local minima, there are two possible
phases and there can be a phase interface between them, where the field varies
from one to the other value, illustrated in Figure \ref{fig:wall}.
Plasma particles approach the interface from either side.  If the interface is
not too sharp, particles follow classical trajectories in the wall
background (wave packets are well described by the WKB approximation).  In the
interface rest frame, a particle's energy and transverse
momentum are conserved, so the classical trajectory is given by
\be
\frac{d \vec r}{dt} = \frac{\p}{E} \, , \qquad
\frac{dE^2}{dt} = 0 \; \rightarrow \;
\frac{2p_z dp_z}{dt} = - \frac{dm^2}{dt}
= -\frac{dm^2}{dh} \frac{dh}{dz} \frac{dz}{dt} \,.
\ee
But $dz/dt = p_z/E$, so
\be
\frac{dp_z}{dt} = - \frac{1}{2E} \frac{dm^2}{dh} \frac{dh}{dz} \,.
\ee
Note that $dp_z/dt$ represents a net force on a plasma particle.  The
total force on all plasma particles, per unit surface area of the wall, is
\be
\frac{F}{A} = -\int dz \frac{dh}{dz} \sum_a \frac{dm^2_a}{dh}
\int \frac{d^3 p}{(2\pi)^3 \, 2E_a} f_a(\vec p,z)
\label{eq:Fgeneral}
\ee
with $f_a(\vec p,z)$ the occupancy of species $a$ with momentum $\vec p$ at
distance  $z$ from the wall. $a$ may be bosonic or
fermionic and $f_a$ is in general out of equilibrium. 
This is the force on particles due to the scalar background; there is an equal
and opposite force on the background, which acts as a pressure difference
restraining the motion of the interface.  The wall also feels a force due to
the vacuum potential; $F_{\rm vac}/A = -\Delta \Vvac$, pushing the wall
forward. Therefore the total force pushing the phase interface forward
is $(\Vvac(h_1)-\Vvac(h_2)) - F/A$.

In general the occupancies $f_a(\vec p, z)$ are all out of equilibrium and it is a
nontrivial problem to determine them, simultaneously with the interface
velocity, shape, and tension.  This is what previous literature has attempted
to do \cite{Turok,Dine,Khlebnikov,Arnold,MooreProkopec,John,Moore}.
However, two limits are very simple.

The first simple limit is 
a bubble wall at rest in the plasma frame \cite{Arnold}.  In this
case, energy conservation, together with Liouville's theorem, imply that $f_a$
remains always in equilibrium.  Therefore
\Eq{eq:Fgeneral} becomes
\be
\frac{F}{A} = 
 \int_{h_1}^{h_2} dh \sum_a \frac{dm^2_a}{dh}
\int \frac{d^3 p}{(2\pi)^3 2E_{p,h,a}} f_a(E_{p,h,a}) \,,
\ee
with $f_a$ the equilibrium distribution function.  This expression coincides
with the expression we found earlier for the thermal potential difference
between phases
$\VT(h_2)-\VT(h_1)$, so the total force on the wall is
$V(h_1)-V(h_2)$.  The critical temperature $\Tc$ is defined as the temperature
where the potential is equal in the two minima.  Therefore, at $\Tc$ the
forces are in balance and a wall at rest will remain at rest, as expected.

The other simple limit is the limit we need in this paper; a wall with
velocity $ v $ which is
ultra-relativistic in the plasma frame, {\em  i.e.}, with gamma factor
$\gamma  \gg 1$.  In practice we will be interested in really
macroscopic values $\gamma \sim 10^9$, so in the following we make no
qualms about strictly expanding to lowest nontrivial order in $\gamma$.
We begin by analyzing what happens in the rest-frame of the bubble wall.
In this frame the plasma is ultra-relativistic with velocity $v$, which
we take to point along the $-z$ direction
(so the plasma-frame wall has positive $v_z$ velocity).
The mean wall-frame particle energy is $E \sim \gamma T$, which will always be
enough%
\footnote{
    In general no excitation should be treated as perfectly massless.
    Therefore, for $\gamma \gg T/m$, the statements here have only
    exponentially small corrections, $\sim \exp(-\gamma m/T)$, as opposed to
    the power (phase space) suppressed corrections one would expect if $m=0$
    strictly.
    }
to allow the particle to pass over the wall, $\gamma T \gg m_a(h_2)$.
Furthermore, the mean momentum $p_z \sim \gamma T$ is so large that the WKB
approximation is excellent and the reflection coefficients are exponentially
suppressed.  Interactions or scatterings between plasma particles
occur at a time-dilated rate $t \sim 1/\gamma T$ (times inverse powers of
couplings) and can be neglected.  Therefore the occupancies evolve
undisturbed.  Also, the number of particles approaching the wall from
the positive $z$ side is exponentially suppressed by $\exp(-\gamma m_a/T)$.
That means that the particles approaching the wall have received no
signal that the wall is approaching, and are in equilibrium.

Rather than integrate the pressure per length over the wall, we can directly
compute the net pressure on the wall by summing up the momentum change of each
particle which passes over the wall \cite{Arnold}:
\be
-\frac{F}{A} = \sum_a \int \frac{d^3 p}{(2\pi)^3} f_a(p,{\rm  in}) 
    \Big( p_{z,\rm in} - p_{z,\rm out} \Big) \,.
\label{eq:F2}
\ee
Here $p$, $f_a(p,{\rm in})$ are the {\em incoming} momentum and occupancy.%
\footnote{
  The integration includes backwards-pointing $p_z<0$ but they will have
  exponentially small $f(p)$ and so the error from their inclusion is
  negligible; the same applies for $p_z$ which are positive but so small that
  the particle would reflect.  
  }
\Eq{eq:F2} is what we would get by performing the
$z$-integration in \Eq{eq:Fgeneral}, keeping $ f _ a $ constant along a classical
trajectory and taking advantage of the absence of reflections.%

Since the wall is at rest and uniform in the transverse directions, both
energy and $p_\perp$ are conserved.  Together with $\gamma \gg 1$,
this makes evaluating $(p_{z,\rm in}-p_{z,\rm out})$ simple:
\bea
p_{z,\rm in}^2 + m^2(h_1) & = & p_{z,\rm out}^2 + m^2(h_2) \,,
\nonumber \\
p_{z,\rm in} + \frac{m^2(h_1)}{2p_{z,\rm in}} + \OO(m ^ 4 /p^3) & = &
p_{z,\rm out}+ \frac{m^2(h_2)}{2p_{z,\rm out}} + \OO(m ^ 4 /p^3) \,,
\nonumber \\
p_{z,\rm in} - p_{z,\rm out}  & = & -
 \frac{m^2(h_2)-m^2(h_1)}{2E} + \OO(m ^ 2/E^3, p _\perp ^ 2/E ^ 3) \,,
\eea
where in the last line we have used 
$E^{-1}=-p^{-1}_z+\OO(m^2/E^3,p_\perp^2/E^3)$.
The terms $ \sim E ^{ -3 } $  will be $\gamma^{-2}$ suppressed and can be dropped.
Therefore the pressure difference is
\be
\frac{F}{A} = \sum_a (m^2_a(h_2)-m^2_a(h_1)) 
  \int \frac{d^3 p}{(2\pi)^3 2E_{p,h_1,a}} f_a(p,{\rm in}) +\OO(1/\gamma^2)\,.
\label{eq:meanfield3}
\ee
The integration measure and occupancies are covariant%
\footnote{
    Note that $p$, $f(p)$ are defined based on their symmetric phase
    ($z>0$) values.  The expression is covariant only if we identify
    $E=E_{p,h_1,a}$ since that is the energy which satisfies the mass shell
    condition in the symmetric phase, where $p$ is defined.
    }
and can be performed in
any frame; it is most convenient to do so back in the plasma frame.
This shows that \Eq{eq:meanfield3} is identical to \Eq{eq:meanfield2}.

Therefore the backwards pressure on the interface in the limit 
$\gamma \gg 1$ is found by replacing the thermal effective potential
$V=\Vvac+\VT$ with $\tilde{V} = \Vvac+\VT[\mbox{mean field}]$.
This is the main result of this section.%
\footnote
    {
    Some readers might worry that this leading-order analysis may receive
    large higher-loop effects of form $\alpha \ln \gamma$.  In particular,
    since the wall ``sees'' particles with extremely high energies, shouldn't
    it analyze their partonic content?  We believe that the answer is ``no.''
    Partonic content becomes important when a particle is analyzed not with a
    large energy, but with a large available transverse momentum needed to put
    the partonic contents on-shell.  The wall is uniform in the transverse
    direction and so cannot impart any transverse momentum; so the actual
    transverse analysis scale is infrared and partonic content is not probed.
    }


Let us comment quickly on scaling.  The density of particles increases as
$\gamma$, due to Lorentz contraction.  But the mean momentum that each
particle induces, in climbing the wall, goes as $1/\gamma$
(since $\delta p_z \sim m^2/E \sim 1/\gamma$), which explains why there is a
finite large-$\gamma$ limit.  

As a check, we can repeat the calculation
in the plasma frame.  Since the wall passes at
the speed of light and purely in the $+z$ direction, it leaves each particle's
$p_\perp$ and $E-p_z$ unchanged.%
\footnote{$E-p_z$ conservation is the same as energy conservation in
  the wall rest frame.}
However $E^2-p^2$ must change by 
$\Delta m^2$, which uniquely determines the changes to a particle's
4-momentum:
\be
(E,p_z,p_\perp) \rightarrow 
\left( E+\frac{m^2(h_2)-m^2(h_1)}{2(E-p_z)} , 
  p_z+\frac{m^2(h_2)-m^2(h_1)}{2(E-p_z)} ,
  p_\perp \right) \,. 
\ee
In computing the momentum transfer to the phase interface, we must
remember to consider the flux of particles through the wall, not the particle
density.  The flux differs from the particle density by a factor of
$v_{\rm relative} = 1 - v_z = 1 - p_z/E$.  Therefore the
force per unit area is
\bea
\frac{F}{A}  & = & \sum_a \int \frac{d^3 p}{(2\pi)^3} f _ a(E_{\rm in}) \:
v_{\rm relative} \: \frac{m_a^2(h_2) - m^2_a(h_1)}{2(E-p_z)}
\nonumber \\
& = & \sum_a \Big( m_a^2(h_2) - m_a^2(h_1) \Big)
\int \frac{d^3 p}{(2\pi)^3} \, f _ a(E_{\rm in})\;
\frac{E-p_z}{E} \frac{1}{2(E-p_z)}
\eea
which is the same as \Eq{eq:meanfield3}.

\section{Example:  Singlet extension of the Standard Model}
\label{sec:analysis}

Here we apply the above results to the simplest extension of the Standard
Model which can provide a strong electroweak phase transition; the
model extended by one real singlet scalar field $s$.
This field need not have an $s\leftrightarrow -s$ discrete symmetry.  If not,
then $s^3$ and $s H^\dagger H$ potential terms can appear at tree level,
potentially leading to a strongly first order electroweak phase transition, as
recently studied in \cite{Profumo}.%
\footnote{Actually the transition can be strong even without these cubic
  terms; furthermore, in the theory with $s\leftrightarrow -s$ discrete
  symmetry, the $s$ particles are an interesting dark matter candidate
  \cite{Pospelov}.}

So consider the scalar sector of an extension to the Standard Model; besides
the complex doublet Higgs field $H$ there is a real singlet field $s$, with
tree-level potential
\begin{eqnarray} 
  V 
  &=&  V _ { H } + V _ { HS } + V _ S
\label{eq:V}
\end{eqnarray} 
with 
\begin{eqnarray} 
   V_{H} &=&  -\mu_h^2 H ^\dagger H + \lambda_h ( H ^\dagger H )^ 2 
   \label{VH}
\\
   V_{HS} &=& \frac{a}{2} H^\dagger H s + \frac{\lambda}{2}
   H^\dagger H s^2 
\\
   V_S &=& \frac{-\mu _s^2}{2} s^2 + \frac{b}{3} s^3 +
   \frac{\lambda_s}{4} s^4 \,.
\end{eqnarray} 
We want to probe whether or not the bubble wall runs away in this
model as a function of the parameters
$\lambda,\lambda_h,\lambda_s,\mu_s^2,\mu_h^2,a,b$, and in particular
we want to see how the wall velocity is correlated with the strength
of the transition (and possibly other attributes).

To do so we perform a Monte-Carlo sampling over the values of the 7 scalar
Lagrangian parameters, fixing $\mu_h^2/\lambda_h$ to give the right
physical Higgs VEV.  For each choice of scalar parameters, we take a
series of steps:

\begin{enumerate}
\item
First we must make sure that the parameters give a vacuum consistent
with phenomenology.  We enforce absolute stability of the theory,
requiring $\lambda_h>0$, $\lambda_s > 0$, and 
$\lambda > -\sqrt{4\lambda_h \lambda_s}$.  Then
we require that the Higgs vacuum is deeper than
any $s$-only vacua and that the scalar states in the Higgs minimum
obey experimental Higgs mass constraints \cite{Higgslimit}%
\footnote{
    We impose a simplified version of the limits found in
    \cite{Higgslimit}.  First one computes the masses of the two
    scalar states and their mixing angles with the pure Higgs
    direction.  Then we approximate the constraint found in
    \cite{Higgslimit} to require $\cos^2\theta<0.25$ if $m<115$GeV and
    $\cos^2\theta < 0.04$ if $m<90$GeV.}.

\item
We compute the thermal effective potential and place the following
constraints on the minimum structure as a function of temperature.
The $h$ minimum (by which we mean any minimum with a nonzero condensate
of the Higgs field) must become unstable (become an inflection point
rather than a minimum) at some point as the temperature is raised; we
call this temperature where it suffers spinodal instability
$T_{{\rm spin}1}$.  Then we require that the $s$-minimum which is
stable above $T_{{\rm spin}1}$ should become spinodally unstable as the
temperature is lowered, at some $T_{{\rm spin}2} > 0$.  The latter
condition ensures that the universe does not become trapped in an
electroweak-symmetric phase as it cools.  (The criterion is a little
too strong; we really only need ensure that the nucleation action to
leave the $s$-phase for the $h$-phase is small enough at some
temperature to allow the transition to occur.  Spinodal instability
means the nucleation rate diverges, so it is a sufficient condition for
this to occur.)

We are mostly interested in the behavior around the transition
temperature and in cases which are marginal between runaway and
finite-velocity bubble walls.  Therefore we expect moderate scalar
VEVs $h/T \sim 1$,
which allows the use of the high temperature expansion for the
thermal potential, which we take to be \cite{DolanJackiw,Kirzhnits}
\bea
\VT &=& h^2 T^2 \left( \frac{9g^2 + 3{g'}^2 + 12y_{\rm top}^2
                +24\lambda_h + 2\lambda }{96} \right)
   +s^2 T^2 \left( \frac{2\lambda+3\lambda_s}{24} \right)
\nonumber \\ &&
   +\frac{(a+b) sT^2}{12}
- \frac{h^3 T}{12\pi} \left( \frac{3}{4} g^3 
+ \frac{3}{8} (g^2+g'{}^2)^{3/2} \right)
\,.
\label{our_pot}
\eea
Here $g,g',y_{\rm top}$ are the weak, hypercharge, and top quark
Yukawa couplings respectively; all other terms arise from integrating
out the scalars.  Note that this potential makes a few approximations,
dropping the effects of Debye screening and the cubic terms for scalar
fields.  It is not our intention to perform a state of the art
analysis, and the dropped effects are not too important if the
transition is relatively strong with relatively light scalar masses
near $T_c$, which is the case when the bubble wall is marginal
between finite velocity and runaway with a transition strong enough to
be physically interesting ($h/T>1$).

Since very weak transitions are physically uninteresting, we discard all
transitions with $h(T_{{\rm spin}1})/T_{{\rm spin}1}<0.2$ from further
analysis.

\item
We determine the critical temperature, where the $s$ and $h$ phases
have equal free energy.  Then we find the nucleation temperature
$T_{\rm nuc}$, which we take to be the temperature where the energy of
a critical bubble is $S_{\rm nuc} = E/T = 100$.%
\footnote{
    It is often stated that bubble nucleation occurs when 
    $\exp(-S_{\rm nuc})= (T/H)^4$ so there is one bubble nucleation
    per Hubble volume per Hubble time.  This is wrong; one should
    define $\Gamma_{\rm nuc} = -dS_{\rm nuc}/dt = H TdS_{\rm nuc}/dT$,
    which is the time it takes for the nucleation rate to change
    appreciably; then one demands that $\exp(-S_{\rm nuc}) =
    (T/\Gamma)^4$ so there are enough bubbles to fill the universe in
    the time it takes for the nucleation rate to change appreciably.
    This guided our choice of 100 (see also \cite{MooreProkopec}).}
The definition of the critical bubble energy is standard
\cite{bubble_nuc} and we use the algorithm for multi-field bubble
finding presented Appendix 2 of \cite{ClineMooreServant}.

\item
Using the potential at $T_{\rm nuc}$ and the procedure of the last
section, we determine whether or not the bubble wall can run away.  In
detail, we must determine two ``potentials'' at $T_{\rm nuc}$; the
actual potential $V(T_{\rm nuc}) = \Vvac + \VT(T_{\rm nuc})$ and the
mean-field potential expanded about the $s$ minimum using
\Eq{our_pot}, $\tilde{V}$.  For our case and thanks to our neglect of
cubic terms arising from the scalar fields, $\tilde{V}(h,s)$ is the same
as \Eq{our_pot} without the $h^3 T/12\pi$ type terms.

\begin{figure}
\centerline{ \putbox{0.45}{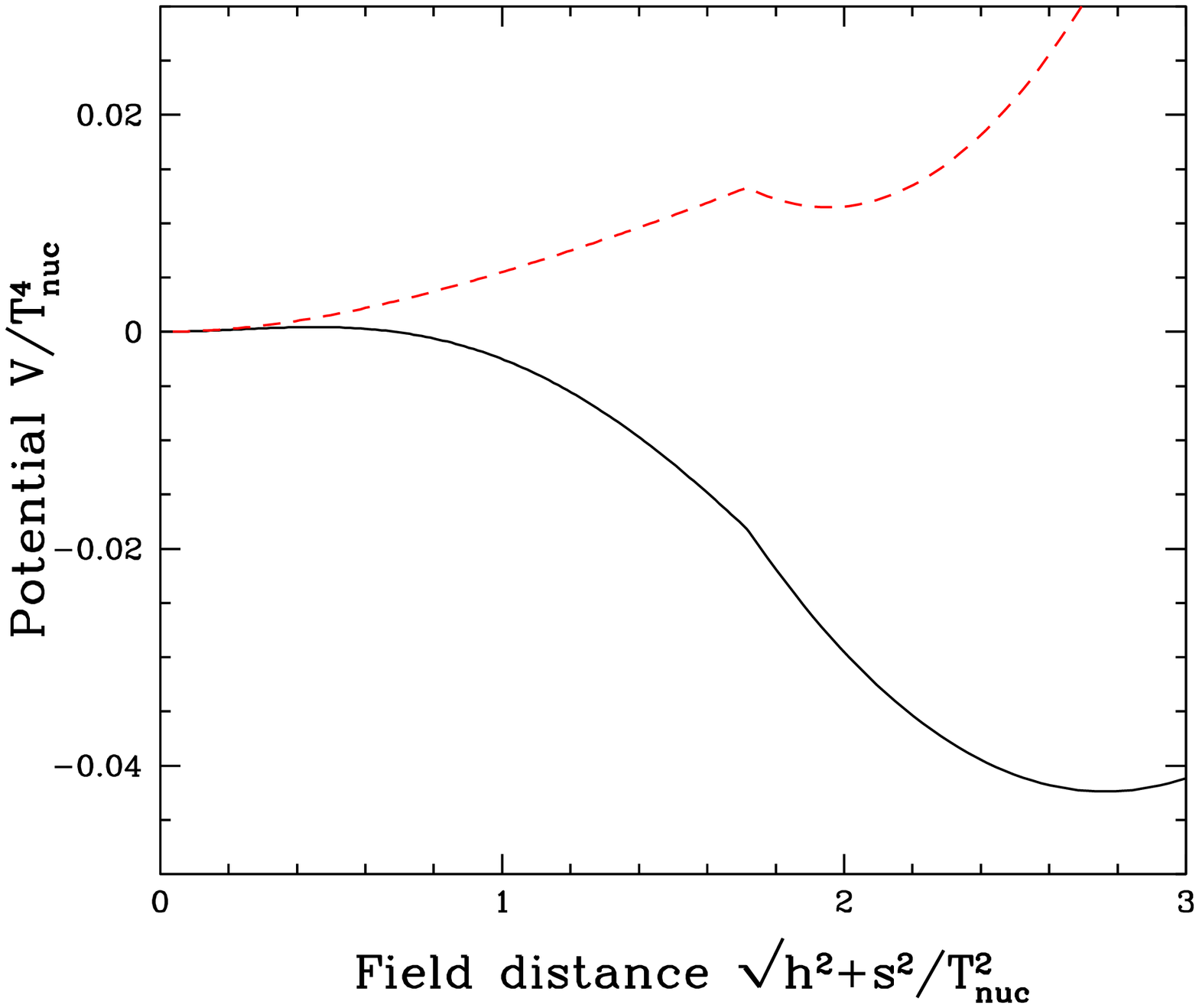} \hfill \putbox{0.45}{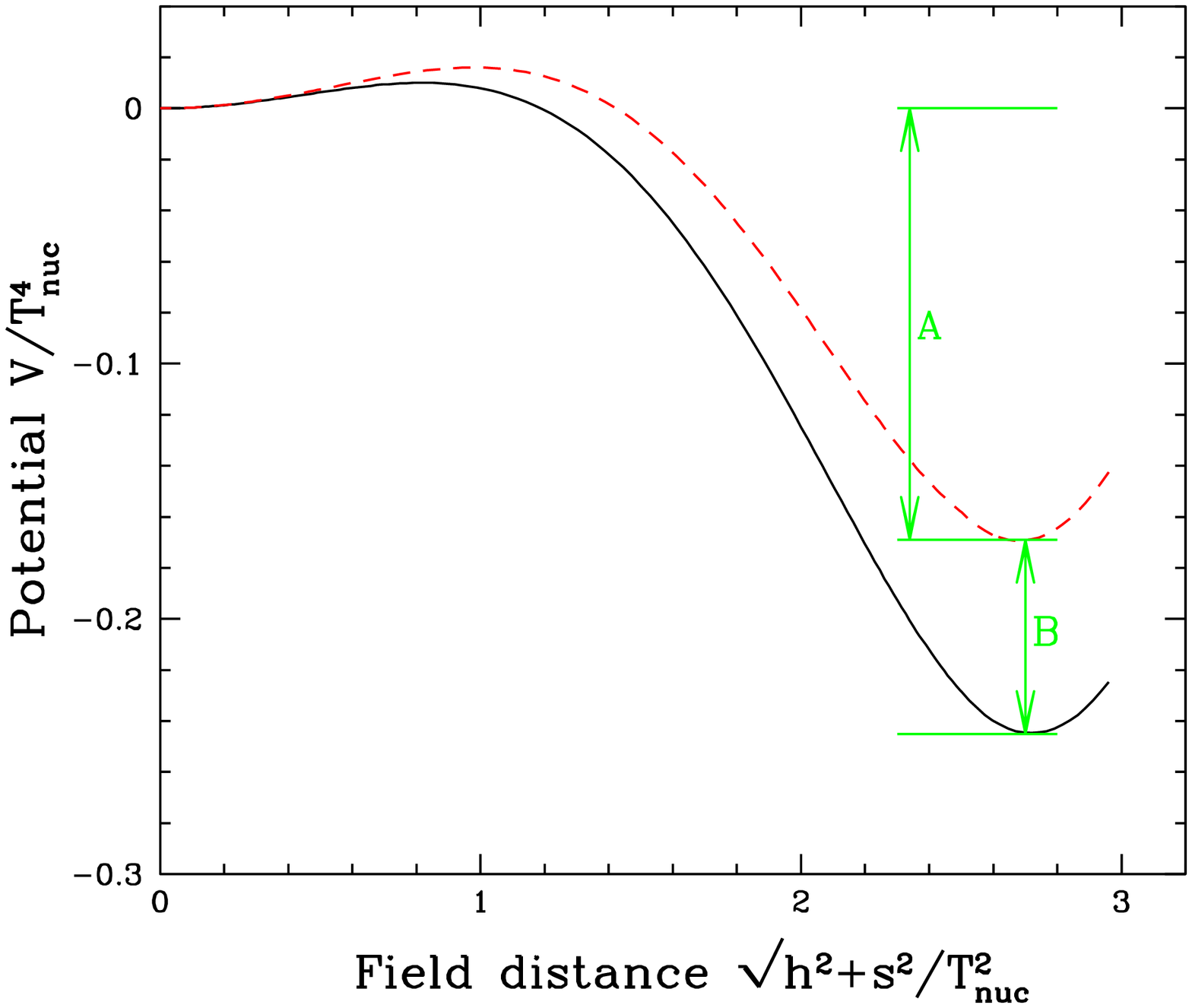}}
\caption{ \label{fig:illustrative} \footnotesize
Example of the effective potential $V$ and pseudo-potential $\tilde{V}$
used to determine bubble wall runaway, for two sets of parameters which
produce the same value of $h/T_{\rm nuc}$.
The left plot is a case where the bubble wall remains of finite
velocity; in the right case it runs away.  Note the different scales on
the vertical axis; the scalar excitations in the two minima are much
lighter for the left case than for the right.
}
\end{figure}

We plot both the true and mean-field potential for two illustrative
cases in Fig.\ \ref{fig:illustrative}.  The black curves are the
actual thermal (equilibrium) effective potentials, evaluated along the
curve in the $h,s$ plane taken by the critical bubble profile.%
\footnote{
    $V$ is evaluated along the
    path through the $(h,s)$ plane which is actually explored by the
    critical bubble; the $x$-axis is the affine distance in the $(h,s)$
    plane along this curve.  The cusp in the left figure is at the
    $(h,s)$ value at the center of the critical bubble; we used a
    straight-line extension from there to the minimum of the potential,
    so the path in the $(h,s)$ plane had a cusp at this point.}
Naturally, in both cases they indicate that the $h$-phase is deeper
(preferred).  The red curves are the mean-field potentials $\tilde{V}$
defined in the last section, which we can
think of as {\em non-equilibrium} effective potentials.  The value of
the red curve at a given value of $h,s$ is the free energy cost per
unit volume to force a phase interface to sweep through the plasma, if
the phase interface changes the plasma from the $h=0$ phase to
a phase with the given $h,s$ value.
For values of $(h,s)$ where it is positive, it would cost
energy to create a phase with these VEVs if the phase is to be
produced by an ultra-relativistic interface.  Where the potential is
negative, there is leftover free energy available from creating the
$h$-phase, which goes into accelerating the bubble wall.  Therefore
the left figure represents a case where the bubble wall cannot run
away.  The right figure represents a case where it will run away.
Note however that the value of $h$ where the red curve is minimized is
not the same as the equilibrium $h$ VEV.  This means that, immediately
behind the bubble wall, the value of $h$ would be slightly lower than in
equilibrium.  The VEV $h$ will relax to the equilibrium value as the
non-equilibrium state after the bubble wall passage re-equilibrates.
The height indicated as $A$ in the diagram is free energy density
available to accelerating the phase interface; the height indicated as
$B$ is free energy density lost to entropy production in
re-equilibrating the system after the wall's passage.
\end{enumerate}

\begin{figure}
\centerline{
 \putbox{0.46}{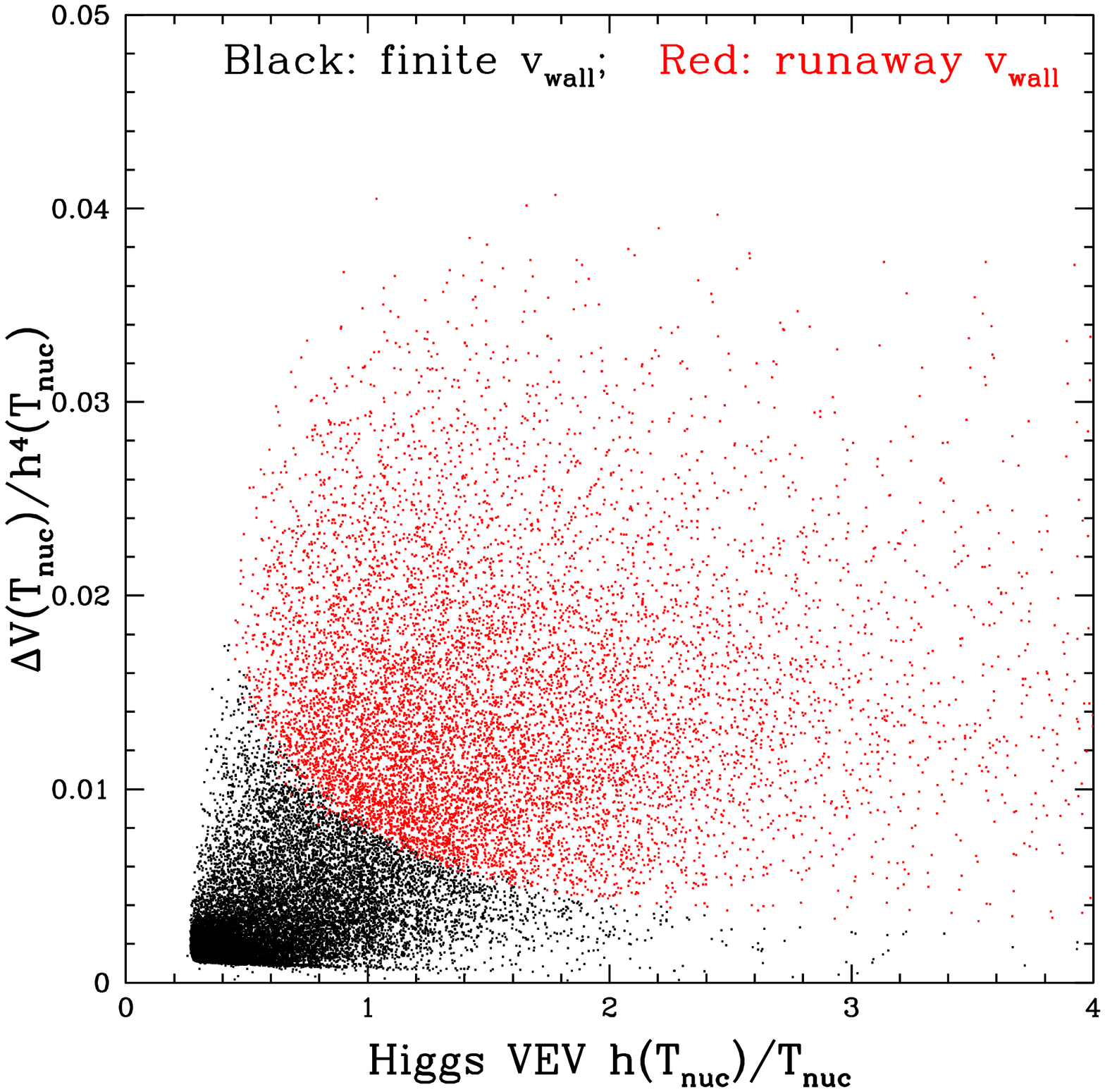}
 \hfill
 \putbox{0.46}{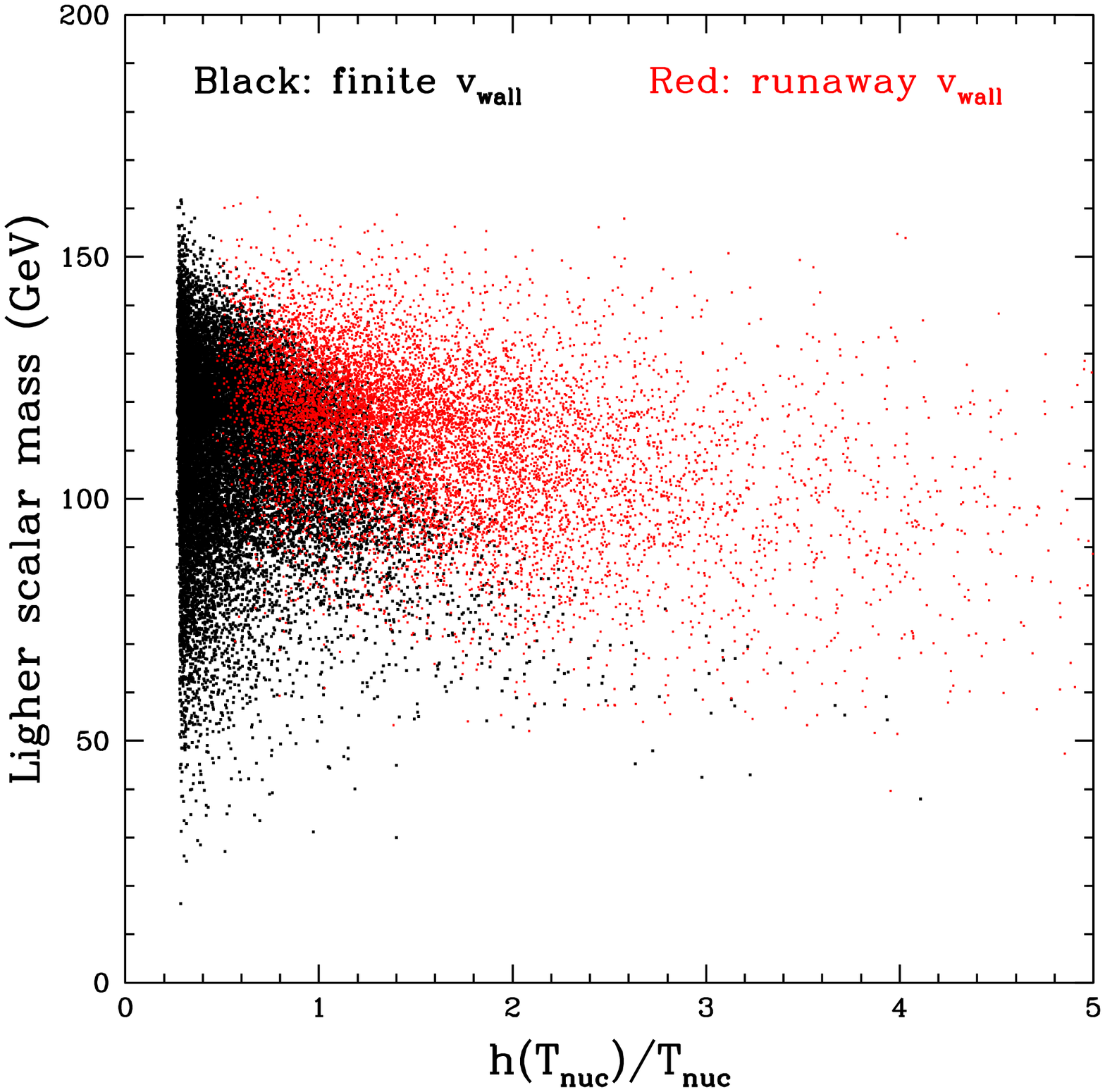}}
\caption{ \label{fig:montecarlo} \footnotesize
  Scatter-plots of $h/T$ versus the free energy available in the
  transition (left) and versus the lighter physical vacuum scalar mass
  (right).  In both figures, red dots are parameter values which give
  runaway bubble wall velocities, while black values give finite bubble
  wall velocities.  For the transition to provide a large free energy or
  to arise from relatively heavy scalars, the bubble wall must generally
  be of the runaway type.}
\end{figure}

Some results from our Monte-Carlo study of the parameter space of this toy
model are shown in Figure \ref{fig:montecarlo}.  The study shows, as
expected, that weak transitions generally involve finite wall
velocities, while strong transitions almost always require runaway
bubble wall velocities.  However a small subset of strong transitions
(as measured by $h/T$, the relevant quantity for baryogenesis since it
determines whether baryon number is violated after the transition
\cite{Baryo-viol}) have finite bubble wall velocities rather than
runaways.  These strong but non-runaway transitions occur when the
potential between the two ($s$ and $h$) minima is very flat.  The figure
also shows that, in these cases,
there is also generally a light physical scalar state
in vacuum, which must be nearly aligned with the $s$ direction in order
to meet Higgs search bounds.  (The cutoff in $h/T$ at small values is
imposed by our procedure since we discard very weak transitions.)

\section{Conclusions}
\label{sec:conclusions}

The physical consequences of the electroweak phase transition depend
on the propagation speed of the phase interface.  Theories with
singlet scalars can have very strong transitions without requiring
large couplings or violating the Higgs mass bound, because the
potential can already possess multiple metastable minima at the
mean-field level.  We show here that such strong mean-field type
transitions generally -- but not always -- involve interfaces which
propagate at virtually the speed of light, as occurs in vacuum
transitions.  However, in the MSSM or other theories where the first
order nature of the transition is fluctuation induced, the bubble wall
can never run away.  We provide a clean, simple criterion, valid at the
1-loop level of analysis, which determines whether or not bubble walls
run away.  Specifically, one defines $\tilde{V}$, the vacuum effective
potential $\Vvac$ plus the quadratic-order Taylor expansion of the
thermal part $\VT$ of the potential about the high-temperature phase.
If the low-temperature phase is preferred at $T_{\rm nuc}$ by
$\tilde{V}$, then the wall runs away; otherwise it has a finite terminal
velocity.  For the case of a real singlet extension of the Standard
Model, we find that very strong phase transitions (say, $h/T>2$)
either have runaway bubble walls or have very slight supercooling, a
small available free energy at the transition, and very light physical
scalar excitations.

It would be interesting to apply this work to the nMSSM.  For cases
where couplings are relatively large or where two-loop thermal effects
are important, it would also be necessary to understand how to
incorporate these effects in to our treatment.

\medskip

\noindent{\bf Acknowledgments}

\medskip

We are indebted to Geraldine Servant, Neil Barnaby and Lev Kofmann for
stimulating us to think about this issue after so many years, and to
Neil Turok, for useful conversations (which occurred about 14 years
ago).  GM would like to thank the Faculty of Physics at
Bielefeld University for hospitality, and
the Alexander von Humboldt Foundation for its support through a F.\ B.\ Bessel
prize.  This work was supported in part by the Natural Sciences and
Engineering Research Council of Canada.

\appendix

\section{Thermodynamics of runaway walls}

\label{the_appendix}

This Appendix explores the thermodynamics which establishes whether
runaway bubble walls are thermodynamically consistent.  Consider two
phases, with pressures $P_1(T)$ and $P_2(T)$; the entropy density
$s=dP/dT$ and energy density $\epsilon=sT-P$ follow from $P(T)$.
Here $P_1(T)$ is for the high temperature $s$-phase, $P_2(T)$ for the
low-temperature $h$-phase, so for $T<T_c$ we have $P_2(T)>P_1(T)$.
The phases are separated by a planar interface moving in the $+z$
direction at the speed of light, with surface energy density
$\sigma(t)$.  We work in the rest frame of the unburnt $s$-phase plasma;
the $h$-phase plasma fills the region $z<t$ and flows with forward
velocity $v _ 2$.  The temperature of the unburnt plasma is $T_1$, the burnt
plasma is $T_2$. We want to determine for what values of $T_1$ the
interface gains surface energy density, $d\sigma/dt > 0$.

This problem has three unknowns; $T_2$, $v _ 2$, and $d\sigma/dt$.  We have
two equations and one inequality with which to determine them;
conservation of energy and of $z$-momentum, and the positivity of
entropy production.  The equation of state provides the stress tensor in
each phase: 
\begin{eqnarray} 
T_i^{\mu\nu} &=&  (\epsilon_i+P_i) u _ i ^\mu u _ i ^\nu + P_i g^{\mu\nu} \, , \qquad
   u _ 1 = ( 1, 0, 0, 0), \quad u _ 2 =(\gamma  _ 2, 0, 0, \gamma _ 2 v _ 2),
   \nonumber \\ 
   &&
   \quad \epsilon_i=\epsilon_i(T _ i), \quad P_i=P_i(T _ i)\,.
\end{eqnarray} 
The index $i$ indicates which phase's equation of state is to be used.
Per unit time, the energy per area of phase 2 increases by
$T_2^{00}$, while the energy per area of phase 1 decreases by
$T_1^{00}$.  These must be balanced by the momentum flows and the
energy change of the bubble wall:
\bea
\frac{d\sigma}{dt} &=& T_1^{00} - T_2^{00} + T_2^{0z} -
T_1^{0z}  \nonumber \\
& = & \epsilon_1 - ( \gamma  _ 2^2 - \gamma _ 2 ^ 2 v _ 2 )\epsilon_2 +
       (1-\gamma _ 2^2 + \gamma _ 2 ^ 2 v _ 2) P_2 \, .
\label{eq:T00}
\eea
Similarly the $z$-momentum must be balanced:
\bea
\frac{d\sigma}{dt} &=& T_1^{z0} - T_2^{z0} + T_2^{zz} -
T_1^{zz}  \nonumber \\
& = & -P_1 + ( \gamma _ 2^2 v _ 2 ^2 - \gamma _ 2 ^ 2 v _ 2  ) \epsilon_2 +
     (1 + \gamma _ 2^2 v _ 2 ^2 - \gamma _ 2 ^2 v _ 2 ) P_2 \,.
\label{eq:Tz0}
\eea
Adding these yields a simpler expression:
\be
2\frac{d\sigma}{dt} = \epsilon_1 - P_1 - \epsilon_2 + P_2 \,.
\label{eq:simple}
\ee

The total entropy of the system must also increase.  The entropy
density of phase 2 is $\gamma _ 2(\epsilon_2+P_2)/T_2$ (the factor $\gamma _ 2$
because of Lorentz contraction) and the volume increase of phase 2 per
unit time is $(1-v _ 2)$.  Therefore entropy increase requires
\be
\gamma _ 2(1-v _ 2) \frac{\epsilon_2(T_2) + P_2(T_2)}{T_2} >
  \frac{\epsilon_1(T_1) + P_1(T_1)}{T_1}
\label{eq:entropy}
\ee
where the RHS is the reduction of entropy stored by the shrinking
$s$-phase.  

The bubble wall runaway is permitted if there is a solution with
$d\sigma/dt > 0$, so energy accumulates in the bubble wall rather than
being lost.  Unfortunately two equations and an inequality are not
enough to solve for the three unknowns $d\sigma/dt$, $v _ 2$, and $T_2$.
If we somehow knew how much entropy is generated at the wall, we could
replace \Eq{eq:entropy} with an equation and all variables would be
determined.  This requires some microscopic dynamics details, which are
treated in the main text of the paper.  Here we will just determine
under what conditions a runaway wall is
thermodynamically consistent.  That is, we can determine the domain of
$T_1$ values where it is at least thermodynamically consistent for a
bubble wall to run away.  We do this by setting the entropy production
to zero, which determines the bubble wall velocity:  treated as an
equality, \Eq{eq:entropy} becomes
\be
v _ 2 = \frac{ 1 - x^2 }{ 1 + x^2 } \, , \qquad 
x \equiv \frac{ (\epsilon_1 + P_1) T_2 }{ (\epsilon_2 + P_2) T_1 }
  = \frac{s_1}{s_2} \,.
\ee
Then the remaining two equations determine $d\sigma/dt$ in terms of the
initial temperature $T_1$.

Specifically, since the $h$-phase becomes more favorable as the
temperature drops, we can search for the temperature at which
ultra-relativistic bubble propagation first becomes thermodynamically
feasible, which is where $d\sigma/dt = 0$ at zero entropy generation.
Then $T_2$ is determined from $T_1$ by using \Eq{eq:simple}:
$\epsilon_2(T_2)-P_2(T_2) = \epsilon_1(T_1) - P_1(T_1)$.  The critical
value of $T_1$ is the value where the so-determined $T_2 $, $ v _ 2$ will give a
valid solution to \Eq{eq:T00}.

We evaluated this value $T_1$ for each set of parameter values in the
toy model discussed in Sec.\ \ref{sec:analysis}.  We also determined
$T_{\rm reheat}$, determined by the criterion
\be
\epsilon_1(T_{\rm reheat}) = \epsilon_2(T_c) \,.
\ee
That is, $T_{\rm reheat}$ is the temperature where the supercooled
$s$-phase has the same energy density as the $h$-phase at $T_c$.  Its
physical importance is that, for slow bubble walls, if 
$T_{\rm nuc} > T_{\rm reheat}$ then the universe will heat back up to
$T_c$ before the phase transition completes and there will be a period
of mixed phase; whereas if $T_{\rm nuc} < T_{\rm reheat}$ then the
supercooling is enough to use up all of the latent heat of the
transition without reheating to $T_c$ and no quasi-equilibrium
mixed-phase situation occurs.

Surprisingly, we find in the toy model that in every case
$T_c > T_1 > T_{\rm reheat}$.  That is, the criterion that
ultra-relativistic bubble walls are thermodynamically consistent is in
practice a {\em weaker} condition than the condition of sufficient
supercooling not to reheat to $T_c$.  Further, we find 
$T_{\rm reheat} > T_{\rm nuc}$ for all but a very small corner of the
parameter space of the theory.  Therefore, in the singlet-scalar model,
supercooling proceeded below either temperature throughout almost the entire
parameter space which provides a first order phase transition.  The fact
that many parameters give bubble walls which cannot run away means that
the production of entropy at the bubble wall significantly reduces the
range in which walls run away, relative to what is thermodynamically
consistent.

\end{document}